# Evaluating Multilingual Metadata Quality in Crossref

Dennis Donathan II*, Mike Nason, Marco Tullney, Julie Shi, and Juan Pablo Alperin*

Correspondence concerning this article should be addressed to Dennis Donathan II at ddonatha@trinity.edu and Juan Pablo Alperin at juan@alperin.ca.

**Author Note**

Dennis Donathan II, Public Knowledge Project & Trinity University.
https://orcid.org/0000-0001-8042-0539

Mike Nason, Public Knowledge Project & University of New Brunswick.
https://orcid.org/0000-0001-5527-8489

Marco Tullney, Public Knowledge Project & Technische Informationsbibliothek (TIB).
https://orcid.org/0000-0002-5111-2788

Julie Shi, Public Knowledge Project & Scholars Portal.
https://orcid.org/0000-0003-1242-1112

Juan Pablo Alperin Public Knowledge Project & ScholCommLab, Simon Fraser University.
https://orcid.org/0000-0002-9344-7439

## Conflict of Interests

We have no conflicts of interests to disclose.

## Funder acknowledgement

This research was supported by Crossref through an award to the ScholCommLab and the Public Knowledge Project. The funder provided technical support for data access, but had no role in the study's design, data analysis, decision to publish, or preparation of the manuscript.

# Abstract

**Introduction**: Scholarly research spans multiple languages, making multilingual metadata crucial for organizing and accessing knowledge across linguistic boundaries. These multilingual metadata already exist and are propagated throughout scholarly publishing infrastructure, but the extent to which they are correctly recorded, or how they affect metadata quality more broadly is little understood.

**Methods**: Our study quantifies the prevalence of multilingual records across a sample of publisher metadata and offers an understanding of their completeness, quality, and alignment with metadata standards. Utilizing the Crossref API to generate a random sample of 519,665 journal article records, we categorize each record into four distinct language types: English monolingual, non-English monolingual, multilingual, and uncategorized. We then investigate the prevalence of programmatically-detectable errors and the prevalence of multilingual records within the sample to determine whether multilingualism influences the quality of article metadata.

**Results**: We find that English-only records are still in the vast majority among metadata found in Crossref, but that, while non-English and multilingual records present unique challenges, they are not a source of significant metadata quality issues and, in few instances, are more complete or correct than English monolingual records.

**Discussion & Conclusion**: Our findings contribute to discussions surrounding multilingualism in scholarly communication, serving as a resource for researchers, publishers, and information professionals seeking to enhance the global dissemination of knowledge and foster inclusivity in the academic landscape.

## Implications for Practice

- Provides insights into the role of metadata quality in determining downstream issues such as discoverability.
- Contributes to the advocacy for multilingualism in academia and scholarly communication infrastructure.
- Academic librarians can apply these findings by encouraging faculty use of persistent identifiers (e.g. ORCiD) to mitigate issues surrounding representation and disambiguation.

# Introduction

Metadata are as vital to scholarly publishing as they are routine. The scholarly community is far more reliant on metadata than most realize; metadata are omnipresent and yet, too often, an afterthought. Found across catalogues, databases, and search engines, metadata are often manipulated to facilitate better discovery, retrieval, and preservation of works.

Metadata also allow for connections between researchers, institutions, organizations, and funding agencies; and between datasets and publications, proceedings, and monographs. Ideally, metadata can also be used to build bridges between different disciplines, regions, and methodologies. In other words, metadata are not just "data about data." Greenberg (2017) suggests that – if implemented correctly – metadata may be better perceived as "value-added language" that "eloquently enables the interplay between an object […] and the desired activity". As such, metadata can expose the narratives of research while also being themselves an essential byproduct of that research.

These functions are thought to be realized when metadata are created and maintained according to shared standards. Standards, however, are subjective and not universally applicable in a global context where scholarly research spans many cultures and languages. While such linguistic and cultural information can be essential markers of value, the prevalence and quality of multilingual metadata in scholarly publishing is not well understood. Similarly, the extent to which individuals and communities are included or excluded from scholarly conversations based on language choices in metadata is unclear. This is in part due to a lack of flexibility within metadata schemas, rules, and systems/platforms to incorporate, advocate for, or, at minimum, acknowledge, the linguistic needs and realities of a global community.

In earlier work, we sought to identify the kinds of metadata issues that can be linked to diverse cultural norms, including publishing by communities and individuals who live and work in languages other than English (Shi et al., 2025). In that work, we identified 32 different types of issues, which we grouped into 5 categories of language, naming, contribution, status, and geography.

Having identified the kinds of metadata issues that could be linked to questions of cultural identity, this study seeks to address further questions about the nature and extent of these issues, especially as they relate to multilingualism. Using a random sample of 519,665 records pulled from the Crossref API, we explore the interplay between language diversity and metadata fidelity. We examine quality through the lenses of completeness and accuracy and discuss the broader sociological, political, and technological implications of multilingualism based on our findings by addressing the following three research questions:

**RQ1:** How widespread are metadata problems in publisher metadata?
**RQ2:** How widespread are multilingual and non-English articles?
**RQ3:** Does multilingualism influence metadata quality?

Our paper begins with an overview of existing literature on metadata quality and multilingualism in research spaces. We then provide an overview of our approach to defining and identifying metadata quality and completeness, followed by our findings for each RQ. After discussing the insights we derived from them, we conclude with the implications of this research for the academic publishing community and possible future directions.

# Literature Review

Strong et al. (1997) define high-quality data as that which "is fit for use by data consumers", an intentionally vague statement that is the foundation from which all remaining frameworks for assessing information and data quality build. Cichy and Rass (2019) evaluate twelve different data quality frameworks and identify the most common attributes: Completeness, Accuracy, Timeliness, Consistency, and Accessibility.
However, Stiller and Király (2017) note that the most common aspect explicitly missing from these frameworks is multilingualism. They concede that multilingualism could be conceived as a "subcategory of other criteria such as completeness, accessibility and consistency", but doing so "risks underestimating the multilingual problem".
Multilingualism as its own dimension of data quality speaks directly to Strong et al.'s (1997) definition by asserting that language is essential to determining if data are fit for use. Indeed, as Russo and Rorissa (2020) note in their discussion on visual resource metadata, multilingualism improves interoperability and reduces language-related ambiguities. Emphasizing multilingualism in science and metadata is therefore key to supporting global scientific collaboration.

## *Multilingualism in Academia*

Language, in particular, has the power to "exclude or include people in conversations and decision-making processes" (Setati, 2008). In the realm of international research and global cooperation, the benefits of using a common language are undeniable. However, the use of English as lingua franca has come with a cost: the diminishment of other languages and the epistemic traditions from which they emerge. This bias for English content is embedded across research tools. For example, popular discovery tools, such as Scopus and the Web of Science, skew heavily in favour of monolingual English articles (Vera-Baceta et al., 2019). This creates a discoverability disparity between researchers who publish in monolingual English journals and those who publish elsewhere. It also perpetuates the concept of "uncitedness", or scholarly outputs that are never cited (Nicolaisen & Frandsen, 2019; Kraker et al., 2021).

Such linguistic biases operate within larger conventional scholarly reward structures that hinge upon the discoverability of research outputs, which depends upon rich, complete, and correct metadata (Achenbach et al., 2022). Quality metadata are, therefore, essential for appropriate attribution, use, and reuse of works. Indeed, analyzing university research data portals, Chiu et al. (2023) found that citation metadata fields were most used among all collected metadata.

Cultural and political tensions underpin these traditional structures. Multiple studies find that these metrics reinforce biases related to status, gender, sex, race, and ethnicity (Nielsen & Anderson, 2021; Davies, et al., 2021; Amutuhaire, 2022). These inequities have significant impacts on the career trajectories of many small and regional publishers and researchers. Research outputs published in languages other than English often receive fewer citations than their English-language counterparts in the same field (Di Bitteti & Ferreras, 2017; Liu, 2017). Moreover, the distribution of non-English is also uneven across disciplines, with non-English found most commonly in the humanities and arts, and English-language outputs dominating in the physical sciences (Kulczycki et al., 2020).

International efforts have worked to underscore the importance of multilingualism in the modern research landscape. The Helsinki Initiative on Multilingualism in Scholarly Communication (2019), for example, promotes linguistic diversity in research and the sharing of knowledge within and beyond academia. The UNESCO Recommendation on Open Science (2021) complements this perspective, defining open science as an "inclusive construct" that aims to "make multilingual scientific knowledge openly available, accessible, and reusable for everyone."

These recommendations recognize that research and science have both local and international benefits, and call on communities to both preserve and actively promote linguistic diversity. They also align with other significant statements such as The San Francisco Declaration on Research Assessment (Raff, 2013) and the Leiden Manifesto (Hicks, et al., 2015), which advocate for greater emphasis on a diversity of metrics, qualitative indicators, pedagogy, community interaction, and scholarly communications.

A robust multilingual scholarly landscape, however, requires equally robust multilingual metadata. Even when the articles themselves are monolingual, it is increasingly the case that multilingual metadata describing these works is created to aid discoverability. As we surface the greater linguistic diversity in research that continues to be obscured, quality multilingual metadata must also be created to describe and connect these threads.

## *What's in a Name?*

"Open discovery infrastructure" that relies on non-proprietary systems and freely available data stands as a possible remedy against systemic discoverability issues (Kraker et al., 2021). Infrastructure, nevertheless, still requires quality metadata to combat discoverability disparities. Alongside the technical considerations of platforms and systems communicating with one another, and compatibility issues between various metadata schemas, quality multilingual metadata must also contend with questions of how language intersects with cultural identities, knowledge systems and research practices, as well as discovery infrastructure.

Searching in languages that do not use the Roman alphabet, for instance, is known to be challenging. It often requires guidance on how to construct search queries, with variations in spellings, word divisions, and word order (Yale University Library, n.d.). Improving discovery for these works typically requires the item and its metadata be translated into English or transliterated from the original non-Roman script (such as Chinese, Russian, or Sanskrit) into the English-language equivalent. Charles et al. (2017) further note that applying language tags consistently to individual metadata elements, instead of to a whole record, and using translated labels and multilingual vocabularies increases cross-lingual discovery. More work is needed to create appropriate metadata schemas and controlled vocabularies for the languages in question, and ensure the interoperability of these schemas and vocabularies as well as relevant tools and systems.

Without improvements in these areas, meanings and identities will continue to be obscured and erased by the limitations and assumptions of monolingual metadata. Take, for example, a standard contributor name field in the NISO Journal Article Tag Suite, JATS 1.3 (JATS Example 1):

```
<contrib contrib-type="author">
    <name name-style="western">
        <surname>McCrohan</surname>
        <given-names>John</given-names>
    </name>
</contrib>
```

**JATS Example 1:** A standard contributor representation in JATS 1.3

This name format is overwhelmingly common for "person" metadata. But, it also makes cultural assumptions based on the maintainers/creators of these standards. A Japanese author, in the same schema, requires three name elements (JATS Example 2). The first against the *ja-Jpan* ISO designation (spanning Han + Hiragana + Katakana), the second an English transliteration, and the third a Katakana transliteration. Submission forms collecting name information do not often feature these alternatives, even when, as in this case, given/surname meanings are a comparative match.

```
<contrib contrib-type="author">
    <name-alternatives>
        <name name-style="eastern" xml:lang="ja-Japan">
        <surname>園田</surname>
        <given-names>直子</given-names>
        </name>
        <name name-style="western" xml:lang="en">
            <surname>Sonoda</surname>
            <given-names>Naoko</given-names>
        </name>
        <name name-style="eastern" xml:lang="ja-Kana">
        <surname>ソノダ</surname>
        <given-names>ナオコ</given-names>
        </name>
    </name-alternatives>
</contrib>
```
**JATS Example 2:** Alternate name representations for a Japanese author in JATS 1.3

Any given work accommodating more than one language might also need to feature multiple names per author, and, possibly, translations or transliterations for titles, abstracts, affiliations, and other article metadata. In the following example of a Chinese author (JATS Example 3), we see a transliteration being spread across two elements, where that name is perhaps better suited to a single string. In this example, we see a contextual shift in meaning.

```
<contrib contrib-type="author">
    <name-alternatives>
        <name name-style="western">
            <surname>Zhang</surname>
            <given-names>Y. P.</given-names>
        </name>
        <string-name name-style="eastern" xml:lang="zh">张轶泼
</string-name>
    </name-alternatives>
</contrib>
```
**JATS Example 3:** Alternate name representations for a Chinese author in JATS1.3

The above examples illustrate another area of concern in multilingual metadata: author disambiguation. With varying cultural norms on name structures, the lack of standardized best practices for transliterating and translating names is an inherent obstacle to the goals of author disambiguation. For instance, Kim and Cho (2013) show that there are twenty-four different name forms for anglicized Korean names. Consider, again, the importance placed on citations in scholarly publishing. Support for persistent identifiers like ORCiD may mute some of the tensions with name metadata and help mitigate these issues, but citation styles will force a choice.

Additionally, system and platform designs may introduce further inconsistencies. The life sciences and biomedical database PubMed, for instance, parses author names according to western conventions, specifically a last-first-middle form that may require cultural compromise in identity. Moreover, inconsistent encodings for non-ASCII characters such as non-Latin alphabets, accented letters, and symbols can prove problematic (Liu et al., 2014). Technical requirements and implementations by databases, repositories, and aggregators lead to technical implementations from publishers that create metadata issues directly tied to cultural identities (Shi et al., 2025).

Abbreviating forenames further obfuscates author disambiguation attempts. Many models utilize forenames in some capacity to accurately match author names, with each additional character increasing the probability of receiving an accurate match significantly (Kim & Kim, 2020). However, in evaluating automated bibliographic detection and extraction models, such as GROBID (Lopez, 2009), Choi et al. (2023) demonstrated that these models have significantly lower accuracy when confronted with multilingual and non-English data.

In addressing issues of disambiguation, it is also important to consider the completeness of the metadata. Wu and Ding (2013) found that using both co-authorship and affiliation significantly increases the ability to disambiguate author names. However, across all geographic regions, publishers frequently submit incomplete metadata. In a study of metadata for arts and humanities articles in Crossref, Borrego et al. (2023) found that listing an author's affiliation was an infrequent practice, and that English language articles have a much lower rate of containing an abstract than Spanish and Portuguese articles.

Multiple barriers prevent researchers and metadata creators from producing quality multilingual metadata: in addition to creating metadata for low-resource languages, or languages that are "less studied, resource-scarce, [and] less computerized", this work requires significant resources and capacity "in terms of time complexity and accuracy" (Khan et al., 2023). It is critical that we do not separate technical factors from the labour required to develop, implement, and maintain them (Horodyski, 2022), and the added challenges faced by those with fewer resources to do so. Labour also encompasses the intentional efforts and decisions needed to ensure that the form and style that metadata assume reflect the needs and realities of the people and cultures that they seek to serve (Rigby, 2015).

# Methods

## *Gathering and Preparing Data*

We set out to programmatically identify metadata issues within a sample of records pulled from the Crossref REST API. Though Crossref is but one of a number of DOI registration agencies available to journals, it is prominent in scholarly publishing and contained over 150 million metadata records as of November 2023. Because of its standing in modern scholarly publishing, Crossref is dedicated to providing access to—and improving the fidelity of—metadata records to downstream consumers and purveyors of open scholarly infrastructure (e.g. ORCiD, OpenAIRE, OpenAlex). That membership, which spanned 148 countries as of December 2022, has significant multilingual needs (Collins, 2022). To represent the realities of its global community, Crossref accepts metadata registered in the language that each member deems most appropriate for their content (Crossref, 2023). This stands in contrast to the English-language requirements of bibliographic databases, like Scopus (Elsevier, 2023).

Focusing on journal article metadata, we pulled a random sample of 531,800 records from Crossref's open REST API using the python package *habanero* (Chamberlain, 2022). We filtered against the "type" field within the metadata to only pull records with the value "journal-article". We used a random sampling method to get a representative cross-section of the larger Crossref data file.[1] We found that the REST API's standard JSON response only contained one abstract in instances where multiple separate abstracts had been submitted. To remedy this for a more rigorous analysis of the metadata submitted to Crossref, we then queried Crossref's XML API to get a more complete version of each metadata record.

Using the *pandas* python package (The pandas development team, 2022), we cleaned the data by removing duplicate records (n=1,757). Due to the previous work done by Shi et al. (2025), we knew that some records were likely mislabelled as "journal-article" within the metadata field "type". Accounting for this, we used string searches for keywords such as "conference", "errata", and "editorial" in the "title" and "journal title" fields to filter out records that had been clearly mislabelled (n=10,379).  We ended up with 519,665 records for further analysis.

---

[1]Crossref metadata contained more than 103 million records for journal articles on December 15th, 2023. See https://api.crossref.org/types/journal-article/works?rows=0 for up to date numbers.

## Language Detection

We used the *py3langid* python library, a Python 3 fork of *LangID* (Lui & Baldwin, 2012) language detection, due to its accuracy, speed, "off-the-shelf" nature, and its demonstrated use for language detection and filtering (Augustin et al., 2023; Capuano et al., 2021). We used the "abstracts" and "article title" fields to detect the language(s) of the record as these fields contained the most substantive text. This reduced the number of records whose language we could identify (since not all records have abstracts), but provided a more accurate result than using the shorter (but more present) title field. The normalized probability threshold was set to 0.95 to avoid false positives on potentially ambiguous text.

After detecting the language(s) present in each record, we used the journal and article "language" fields and abstract language attributes within the records in conjunction with our detected languages to categorize the records into three distinct groups, or Language Types (LT): Monolingual English, Monolingual Non-English, and Multilingual.

## Metadata Completeness

Before they can be indexed, metadata need to exist. Metadata completeness—the degree to which metadata elements are present or not—is a baseline for effective search and retrieval of scholarly works. While completeness is considered an aspect of metadata quality, it is often mistaken as synonymous with accuracy. Completeness, though, can be a deceiving measure that masks many forms of errors.

In this study, we assessed the completeness of metadata records by examining a subset of metadata fields: "article title", "abstracts", "journal language", "article language", "authors", and the author sub-field of "affiliation". These fields were selected for their representation of research and authors and their influence on discoverability. We consider an absence of values in these fields to be an issue, even when, according to the Crossref metadata schema, only the "article title" field is "required". The "authors" and "abstracts" fields are only "recommended", and "language" fields fall outside the recommendations altogether (Crossref, 2021).

## Metadata Quality

Metadata quality is more than completeness; it also encompasses accuracy, consistency, and adherence to best practices. Using the framework developed in Shi et al. (2025), we investigate 5 dimensions of metadata quality, focusing on issues that have potential for cultural or representative ramifications and records from non-English and multilingual publishers.

As with completeness, we limit our analysis to a few specific metadata fields, including: "article title", "abstracts", "authors" and several of its sub-fields, "given_name", "surname", "sequence", and "affiliation". For the "authors" field, several issues were examined. First, we checked for the presence of affiliations represented as authors instead of in the "affiliation" sub-field. This was accomplished by looking for the subfield "name" in lieu of the traditional structure of "given" and "family". We also examined the absence of any affiliation metadata for all contributors in a given record. Additionally, we inspected the "given_name" and "surname" sub-fields to identify cases where only initials were provided or if non-ASCII characters were present. Lastly, we examined the "sequence" sub-field to determine how many authors of a record were listed as "first" authors.

The "abstracts" field was assessed to determine the presence of multiple languages within a single abstract using the same python library as above, and an issue was recorded when multiple languages were identified in this field. The resulting dataset is available at Donathan et al. (2025).

# Results

## *RQ1. How widespread are metadata problems in publisher metadata?*

Building on our previous exploratory work outlining metadata issue categories (Shi et al. 2025), we analyzed the completeness and quality of the following metadata elements: "article title", "abstracts", "authors", "journal language", and "affiliation".

### *Sample Completeness*

Completeness is the easier of the two metadata problems to identify programmatically. In *Table 1*, we see the prevalence of missing metadata—alongside their definitions—across the sample as a whole.

**Table 1.** Metadata completeness issues detected programmatically from sampled records.

| Issue | Definition | Prevalence |
|---|---|---|
| Article language absent | Value is not provided for the language of an article. | 95.7% |
| Abstract absent[2] | The abstract of the item is not provided within metadata. | 75.8% |

---

[2] This measured value is slightly underestimated, since some publishers fill this field with placeholder text instead of leaving it empty. Examples include *<jats:p>No abstract is available for this article.</jats:p>, <jats:p>Abstract not Available.</jats:p>* and the empty *<jats:p />*.

| Affiliation Absent | There are no affiliations listed for any author within a record | 74.3% |
|---|---|---|
| Journal language absent | Value is not provided for the language of the journal. | 20.9% |
| Author(s) absent | Value is not provided for the author(s) of the article. | 9.6% |
| Title absent | Value is not provided for the title of the article. | 0.3% |

Given our focus on multilingualism, one of the more glaring absences in our sampled records is of metadata for the language the article was written in.[3] While 20.9% of all sampled records (n=108,465) were missing language metadata at the title-level, 95.7% percent (n=497,541) were missing language metadata at the article-level.

Abstract metadata were missing in 75.8% (n=394,117) of the records we sampled. This is significant, but not unexpected. The absence of abstracts in article metadata is a well-documented issue that collaborative projects, such as the *Initiative for Open Abstracts* (2023), are working to address.

A much smaller 9.6% (n=50,089) of the records sampled did not include author metadata. While some of these records refer to news, editorial information, etc., or to documents authored anonymously or by institutions, metadata best practice still calls for providing information about the authors.[4] The Crossref schema accommodates these situations by not requiring author metadata for a valid deposit. Here, we see a clear tension between "completeness" and "accuracy" and are left to make assumptions about the quality of the metadata record.

Article title metadata were missing in 0.3% of all sample records (n=1,473), despite this element being required in the Crossref schema. Further investigation of the related records reveals two primary causes: a valid registration can (or *could*, at some point in Crossref's past) be forced by either placing a single whitespace within a title field or by providing a value in the "original-title" field—usually in a non-English language—while leaving the "title" field blank.

### *Sample Quality*

To evaluate quality, we stayed within the scope of metadata elements chosen for assessing completeness. Quality is more difficult to determine programmatically and, as such, our results are more limited.

---

[3] Note that language information is not mandatory in Crossref metadata.

[4] https://www.crossref.org/documentation/principles-practices/best-practices/bibliographic/#contributors

We paid particular attention to the language metadata, and these results are outlined in *Table 2*. Because we could not rely on the accuracy of language stated in the metadata of the record itself, we ran an algorithm on the combined title and abstract metadata for enough text to accurately detect language. Of records with abstracts, 1.6% used more than one language in the field (this corresponds to 0.4 percent across the entire sample). Typically, this suggests an abstract has been translated into at least one other language, but all translations were improperly recorded in a single field instead of in separate abstract elements, each with a proper language attribute, as recommended by Crossref.[5]

**Table 2.** Language quality and completeness issues detected programmatically from sampled records.

| Issue | Definition | Prevalence of Issue (entire sample) | Prevalence of Issue (only records with corresponding metadata) | Number of records with corresponding metadata |
|---|---|---|---|---|
| Language does not match | The value provided in the journal language field does not match values given in the Title field | 2.5% | 3.1% | 411,200 |
| Multilingual abstract | The provided abstract contains multiple languages within a single field | 0.4% | 1.6% | 125,548 |
| Language does not match | The value provided in the journal language field does not match values given in the Abstract field | 0.6% | 3.6% | 85,118 |

Even when the language attribute of an abstract was recorded in the metadata, we discovered instances where this stated language differed from the language we could detect in the title and abstract fields. Some of the abstracts were, in fact, multilingual (two abstracts merged into one field). Some abstracts were in English, although the articles were not. In other cases, the language detected was not English, but the metadata stated it was.

---

[5] https://www.crossref.org/documentation/principles-practices/best-practices/multi-language/

Quality issues across author metadata were identified and detailed prior in Shi et al. (2025). Different assumptions and practices, as well as discrepancies between user language, software language support, and diverse differences in workflow, may influence the creation and transmission of author-related metadata. *Table 3* lists the issue, definitions, and prevalence of author metadata quality issues. The prevalence of issues is given in relation to the entire sample of 519,665 records as well as in relation to the particular subset of records that contain corresponding data (n=469,576).

**Table 3.** Author issues detected programmatically within both the full sample and a subset of records containing corresponding data.

| Issue | Definition | Prevalence of Issue (entire sample) | Prevalence of Issue (only records with corresponding metadata) |
|---|---|---|---|
| Organizations presented as authors | An organization is listed as an author instead of being placed in the appropriate affiliation field | 1.9% | 2.1% |
| Use of honorific in name | A professional title or honorific is included in the author's name. | 0.1% | 0.1% |
| Name in all capital letters | An author's name, either given or family, is input in all capital letters | 4.9% | 5.4% |
| All authors listed as "first" | All authors within a record and all have been noted as "first" within the sequence field | 1.3% | 1.4% |
| No first author identified | No authors within a record have been noted as "first", instead all listed as | 0.3% | 0.3% |
| | "additional" within the sequence field | | |
| Only initials provided | Author names, either in the given or family name fields, are given only as initials, such as X.Y, or X Y or XY, etc. | 22.8% | 25.2% |

Instances where author names are provided solely using initials is the most prominent metadata quality issue across all records (22.8%, n=118,397). These are cases in which at least one author name in a record is provided using only initials. This amounts to a quarter (25.2%) of all articles that contain author metadata.

For 1.9 percent of all articles (n=9,609; 2.1 percent of articles with at least one author), we discovered institutions recorded as "persons" in the record. Crossref's schema does allow for organizations (as well as anonymous authors) as contributors in a separate metadata element. In the cases we detected, institutional names were incorrectly recorded as "person" contributor metadata. Less frequent (0.1%, n=441) was the use of honorifics like "professor" or "dr." in the author name metadata elements.

In some cases, author names are provided in all capital letters. As discussed in Shi et al., (2025), capitalization, especially when not all authors within a journal or within a record are written in the same form, can be indicative of users attempting to code additional information, such as the individual's status or position. In our sample, we found all-caps author names in 4.9% of all records (n=25,311), and 5.4% of all records that contain author metadata.

Crossref allows for information on author sequence, with authors optionally being labelled as "first" or "additional." The overwhelming majority (99.7%) of all sampled records contain at least one author marked as "first" (including sole-authored papers). A much smaller proportion (2.4%) of sampled records have multiple authors labelled as "first." In the majority of these cases (1.3% of sampled records) have all of the authors labeled as "first". This latter group is potentially problematic, as it could signal that the label was indiscriminately applied to all authors.

## RQ2. How widespread are non-English records and multilingual records?

Looking at the subset of records with abstracts (n=125,548), we determined the languages of abstracts by combining the explicit language attributes from Crossref's XML API with results from the language detection algorithm, creating a unique list for each record. Ultimately, we identified 81 unique languages within the sample. English was observed most frequently, being present in 90.6% (113,697) of records, followed by

Portuguese (3.88%, 4,865), Spanish (3.21%, 4,033), French (2.84%, 3,567), and Malay (2.55%, 3,204).

Some abstracts contained more than one language. Utilizing the various language metadata fields and the languages detected in both the abstracts and the titles, we were able to group the records into three distinct groups: Monolingual English, Monolingual Non-English, and Multilingual. We were unable to classify 3.1% (15,888) of records, which we group under "Uncategorized." Records in this group did not state a language for the journal or the article, and either did not have enough available text for the language detection to reliably detect the languages used, or contained several ambiguous terms, such as loan words that are incorporated from one language into another and technical or scientific terminology (ex. *N-Pivaloyl-N´-dichlorophenyl*; *Clematis vitalba*), that prevented the language detection algorithm from returning a result that exceeded the .99 normalized probability threshold. *Table 4* shows the number of records that are assigned to each of the Language Type groups.

**Table 4.** Language Type groups and their size according to number of records and publishers.

| Language Type | Number of Records | Percent of Total (%) | Number of Publishers | Mean Records per Publisher | Number of Journals |
|---|---|---|---|---|---|
| Monolingual English | 439,231 | 84.5 | 8,614 | 51 | 43,036 |
| Monolingual Non-English | 38,810 | 7.5 | 4,023 | 9.6 | 13,171 |
| Multilingual Record | 25,736 | 5.0 | 2,940 | 8.8 | 8,939 |
| Uncategorized | 15,888 | 3.1 | 2,271 | 7 | 6,120 |

Of the 25,736 Multilingual records, 24,242 (94.2%) use English as one of their languages, followed by German (30.6%, 7874), French (17.9%, 4609)  and Portuguese (8.6%, 2201). In total, we found English was present in 89.2% (463,472) of records.

## RQ3. Does multilingualism influence metadata quality?

Notably, 77.3% (339,525) of Monolingual English records are missing abstract metadata, whereas 47.3% (12,173) of Multilingual records and 69.1% (26,818) of Monolingual non-English records are missing abstracts. Additionally, the differences across categories in the presence of title-level language metadata is considerable, with 55.5% (21,345) of Monolingual non-English and 25.4% (6,511) of Multilingual records missing title-level language metadata as opposed to 15.3% (64,567) of the Monolingual English records not providing title-level language metadata.

 *Figure 1* represents the prevalence of completeness issues by language category and *Figure 2* breaks down the prevalence of quality issues within each language category. There are some issues that are unique to certain categories. For instance, only Multilingual records will have abstracts in more than one language (n = 2,099). Similarly, records that are missing an article title (n = 1,473) are all within the Uncategorized language type because metadata in those records is so sparse that we cannot reliably determine the language of the record.

**Figure 1.** Prevalence of missing values compared between Language Type

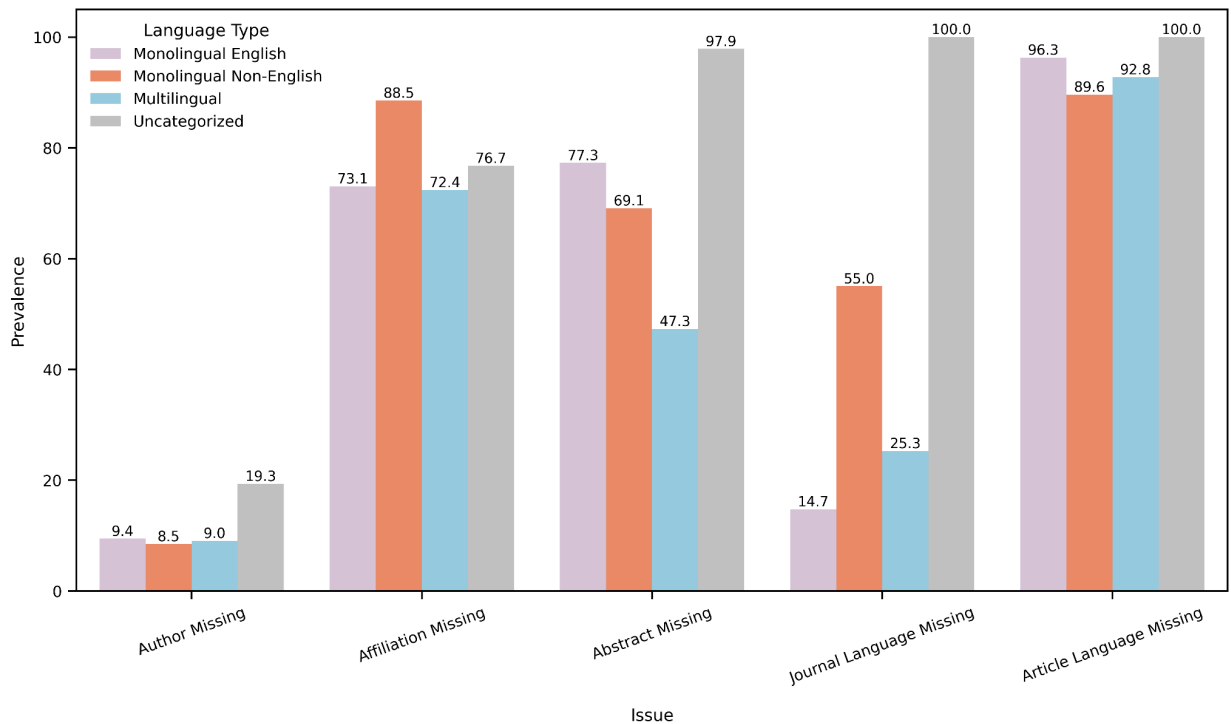

**Figure 2.** Comparing prevalence of metadata quality issues between Language Types

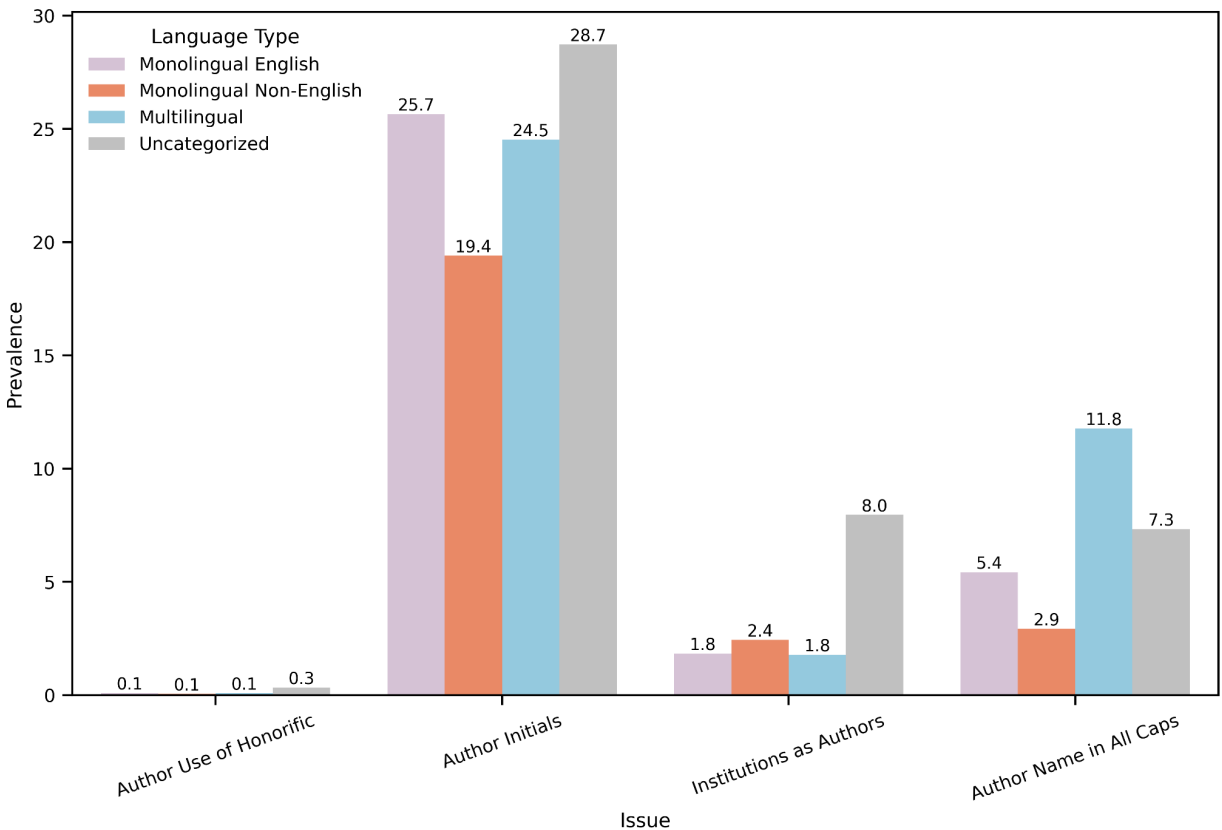

Notably, we found that 28.8% (7,415) of Multilingual records have at least one of the author issues depicted in *Figure 2*. Comparatively, we determined that 27.1% (118,881) of Monolingual English records and 21.6% (8,390) Monolingual non-English records have issues across the same author categories. The most notable difference between LTs can be found in "Author Name in All Caps" as the Multilingual records engage in this practice at a rate more than double that of Monolingual English records.

# Discussion

In this paper, we conducted an analysis of metadata quality, metadata completeness, and multilingualism in Crossref records. We aimed to find out the prevalence of multilingual records and the variances, if any, between the different language types in terms of the completeness and quality of metadata registered with Crossref. We also discovered myriad limitations inherent in programmatic evaluations of metadata fidelity, as well as limitations in the metadata available in the JSON Crossref API.

Our findings reveal several important insights. We found that English continues to dominate academic publishing among Crossref members (84.5% of records in our sample were exclusively English, and 94.2% of multilingual records included English metadata). Overall, English was present in 89.2% of all records we sampled. These proportions are in dramatic contrast to similar data found for journals using the open-source publishing platform, Open Journal Systems (OJS) (Khanna et al., 2022). In that study, Khanna and colleagues sampled 996,000 documents from journals using OJS and found that only 55% of records could only be found in English (Indonesian was the next most common additional language). In contrast, in our sample, we found 12.4% of records contained at least one non-English language, most commonly German, French, or Portuguese.

It is possible that the overwhelming majority of English may be seen as both an explanation and an excuse for the embedding of metadata norms that presume English. However, as the number of non-English records (both among Crossref members and users of OJS) attest, there is a latent and growing need to recognize and value multilingualism in scholarly communication, particularly through metadata. We are, though, limited by what we can know about the provenance of an article based exclusively on metadata. An English-language submission platform may default to English language metadata attributes (or not populate these metadata at all) and be easily overlooked by authors. Platforms with intentional tools for multilingual content may improve uptake for multilingual publications. Or, alternatively or in addition, it is possible that the kinds of independent publications using OJS are more likely to be multilingual. This warrants further investigation.

What our data—or the lack thereof—tells us about completeness is more actionable. While the proportion of missing abstracts on the whole is high (75.8% in our sample), there is reason to believe that efforts such as the Initiative for Open Abstracts (I4OA) are successful (Van Eck/Waltman, 2021). An analysis of publications reveals that only 48% of the two most recent years of publications are missing abstracts (Kramer, 2024). It may be that the remaining gap can also be addressed through alternate means. Kramer's (2024) analysis of abstracts available through OpenAlex showed that they were able to reduce the number of Crossref DOIs with missing abstracts to just 13%, for the most recent two years, through web scraping. Crucially, however, we found that there is a difference of 30 percentage points in the prevalence of abstracts between multilingual records and monolingual English records. We have also found a higher share of available abstracts for monolingual non-English records compared to monolingual English records which supports a previous analysis (Borrego et al., 2023). Abstracts provide valuable information about the content of the publication and should be encouraged in all languages, but our analysis suggests that multilingual and non-English monolingual journals are putting more effort into leveraging the benefits of abstracts as a vital part of bibliographic discovery.

Similarly, a very high proportion of records (74.3%) were missing affiliation data entirely, which lines up with what was found by Zhang et al. (2024). However, unlike abstracts, non-English monolingual and multilingual records were less likely to contain affiliation metadata when compared to monolingual, English records. Affiliation data is known to be missing, or partially missing, as was confirmed by a recent study using OpenAlex data (Zhang et al., 2024). However, this metadata is crucial for any institutional analysis, as well as for analyses that require author name disambiguation (Wu & Ding, 2013). Name disambiguation may also be challenged by the prevalence of author names presented solely as initials (22.8% of the analyzed records). However, unlike abstracts and affiliations, data quality in name fields varies little by language.

While in some cases, such as with abstracts, the missingness may be due to publisher policies (Kramer, 2024), many missing or erroneous metadata can be either a product of inflexibility of the publishing infrastructure (e.g., the manuscript submission system or the metadata exchange formats) or a misunderstanding of this metadata by authors or editors. Further work is needed to better understand the origins of the types of errors we found.

There is a complicating factor in considering the completeness or accuracy of multilingual metadata available in Crossref. One of the more dramatic absences in record metadata was an indicator of language for both journal language (title-level) and article language.

From a flexibility standpoint, multilingual journals are immediately limited by a single language attribute for title-level metadata. Titles publishing in multiple languages are obligated to pick a primary one. We observed an elevated number of missing journal language metadata for records labelled as either non-English monolingual or multilingual. Given requirements from indexing services, journals prioritizing discovery may opt to select or presume English even when it isn't the primary language. Or, potentially, the same presumption of English could result in missing language attributes entirely.

Crossref also supports language attributes for article and abstract metadata. It is in the discrepancies, obfuscation, and omissions of these metadata that a broader picture of multilingual metadata in journals becomes especially muddy. 95.7% of sampled records didn't have an article language attribute at all, so title-level metadata is especially crucial. Multilingual abstract metadata is dependent on how the metadata is retrieved. Crossref supports multilingual metadata in deposits, and full metadata records are retrievable via their XML API. Here, you may see multiple abstracts with associated language attributes for every available language. But, when the data is pulled from the more feature-rich and public-facing REST API, only the first listed abstract is pushed downstream to users.

Crossref is transparent about this limitation in their documentation, but it could have any number of possible implications. For example, journals publishing works in many languages may front-load English abstracts and choose English as their primary language for the purpose of indexing, even if most of their articles aren't written in English. It also means that users of Crossref metadata are likely not being exposed to portions of the multilingual metadata that *does* exist. Discovery platforms like OpenAIRE or OpenAlex, for example, use the REST API. Multilingual publications and authors may well be performing more labour to record accurate metadata, but the value of that work is not exposed to services downstream.

This stress is further exemplified by the rates at which empty language attributes occur across language types: 25.3 percent of multilingual and 55 percent of non-English monolingual records did not record a journal-level language attribute, as compared to 14.7 percent of monolingual English records. Given the notion of English as lingua franca in science, one might expect the assumed default to translate to more missing language attributes in English records. Our results indicate that this is not the case.

These problems have a compounding effect on one another. Missing language metadata necessitates detecting a language programmatically. Language detection works best on larger bodies of text, but a significant number of publications did not provide abstract metadata at all. When they did, the nearly total absence of article-level language metadata, contradictions between declared languages (when they are provided) and the language of the abstract, and the ambiguity involved in submitting multiple abstracts for a single record might not instill any confidence in language metadata. Affiliations might provide additional context clues for publications, but those are also unreliably available in Crossref data. The result is that it is difficult to see why these issues happen or to draw clear conclusions about intent. This, in turn, leads to a long series of open questions: Are journals gaming metadata to improve indexing? Are publishers not recording this metadata themselves or, more likely, not *providing* this metadata to Crossref? Why? Would providing all abstracts in all languages improve language reporting broadly? Could downstream consumers of Crossref data support that?

## Conclusion

Metadata problems are widespread in scholarly publishing. One might assume that because records with multilingual metadata require more content, with more variability, qualifying attributes, and in more fields, that these records would have a higher likelihood for error. More metadata; more problems. In this study, our findings show that the introduction of multiple languages to metadata records deposited to Crossref do add more problems, but not to a substantial degree, and not for all types of errors.